\documentclass[a4paper]{article}
\pdfoutput=1
\usepackage{color}
\usepackage{soul}
\usepackage{graphicx}
\usepackage{amsmath,amsbsy,amsfonts,amssymb,mathrsfs}
\usepackage[version=3]{mhchem}
\usepackage{enumerate}
\usepackage{textcomp}
\usepackage[utf8x]{inputenc}
\usepackage{float}
\floatstyle{plain}
\newfloat{scheme}{htbp}{los}
\floatname{scheme}{Scheme}

\usepackage[sort&compress]{natbib}
\bibpunct{[}{]}{,}{s}{}{}

\soulregister\cite{7}

\newcommand{\fig}{Figure}
\newcommand{\Fig}{Figure}

\newcommand{\figref}[1]{\fig~\ref{#1}}
\newcommand{\Figref}[1]{\Fig~\ref{#1}}

\newcommand{\schemeref}[1]{Scheme~\ref{#1}}

\renewcommand{\eqref}[1]{Eq.~(\ref{#1})}

\title{Atom-Tunneling in Chemistry}

\author{Jan Meisner, Johannes K\"astner$^*$}

\date{}

\usepackage{setspace}    
\AtBeginDocument{\doublespacing}

\begin{document}

\maketitle

\noindent
\includegraphics[width=15cm]{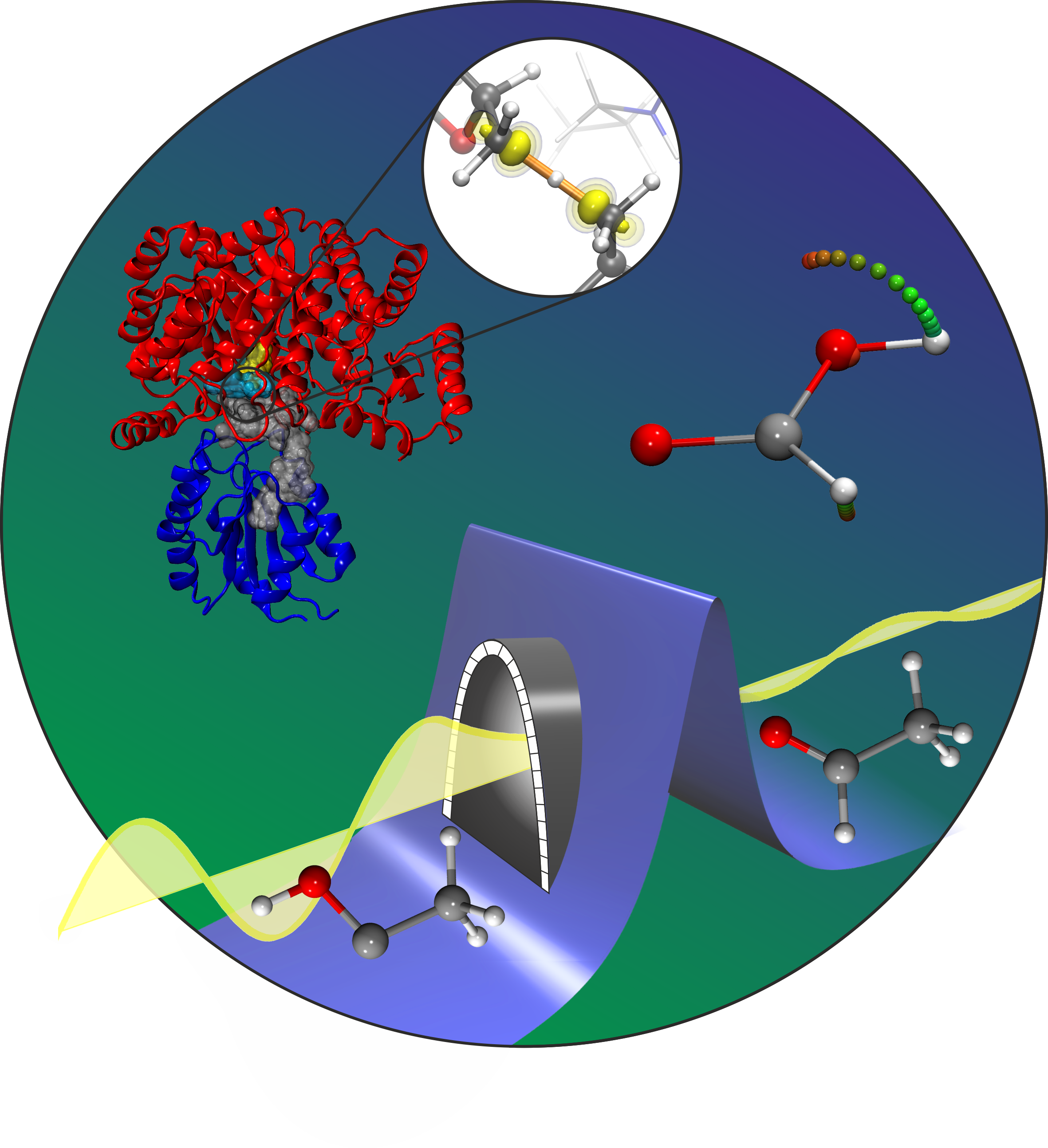}



\newpage

\noindent $^*$ Jan Meisner, Prof. Dr. J. Kästner\\ Institute for Theoretical
Chemistry University of Stuttgart Pfaffenwaldring 55, 70569 Stuttgart
(Germany) E-mail: kaestner@theochem.uni-stuttgart.de

\bigskip

\noindent
\begin{minipage}{10.cm}\small
\textbf{Jan Meisner} (born in 1988)
obtained his master's degree in chemistry from the University of
Stuttgart. After research visits at the University of Heidelberg and at
Imperial College, London he started his doctorate with Johannes K\"astner.  He
studies the tunnel effect in chemical reactions relevant for astrochemical
networks.
\end{minipage}
\begin{minipage}{3.3cm}
\includegraphics[width=3.2cm]{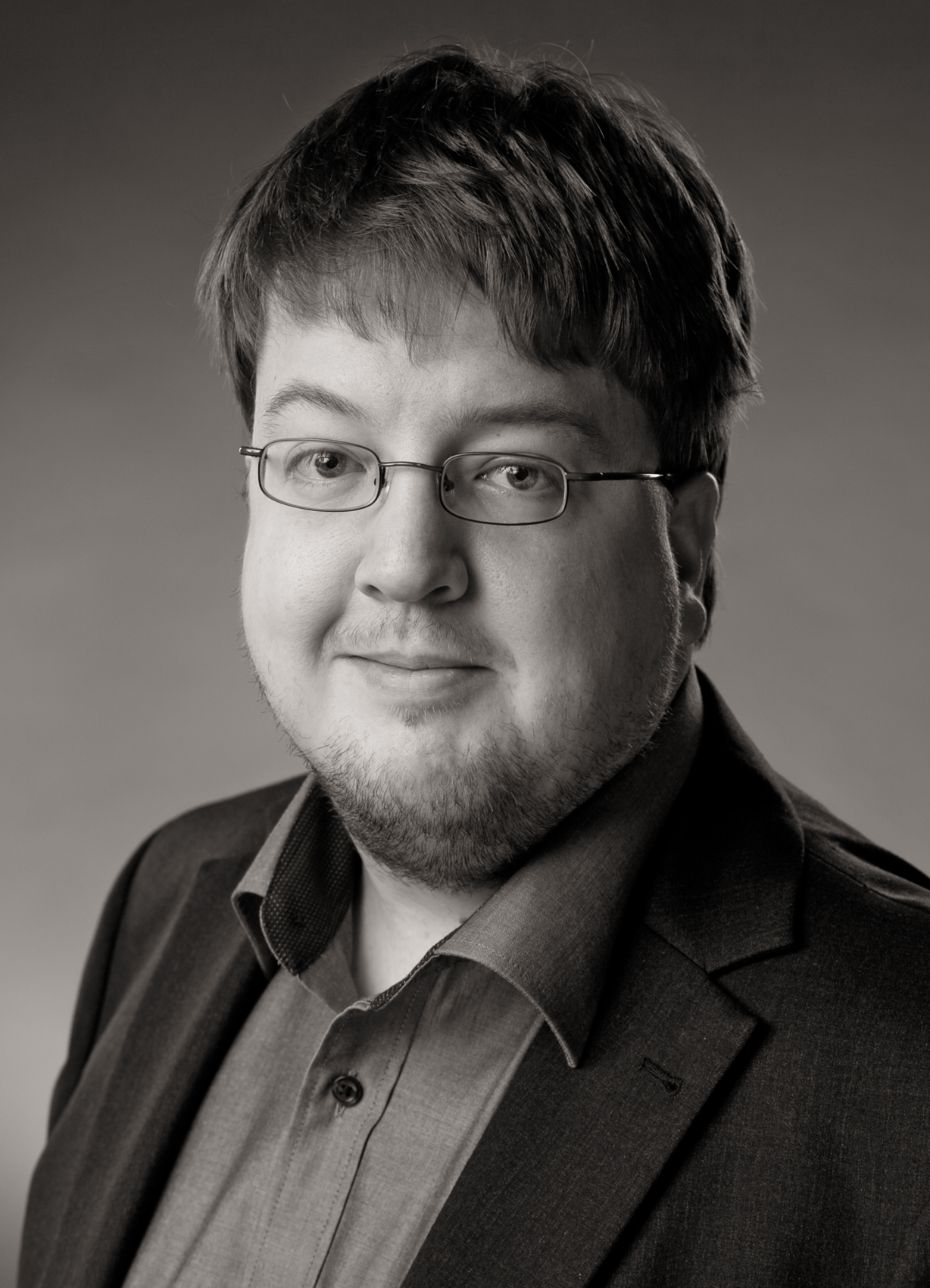}
\end{minipage}

\bigskip

\noindent
\begin{minipage}{10cm}\small
\textbf{Johannes Kästner} received his degree in chemistry from TU Vienna and
his Ph.D. in physics from TU Clausthal. After postdoctoral stays with Walter
Thiel at the Max-Planck Institute in Mülheim and at Daresbury Laboratory, UK
he became Juniorprofessor at the University of Stuttgart, where he was
promoted to ordinary professor in 2014 (tenure-track). His main research areas
stretch from reaction rates and tunneling phenomena to enzymatic mechanisms
and free-energy simulations. Johannes Kästner received the Hellmann prize in
2012 and the OYGA Award in 2015.
\end{minipage}
\begin{minipage}{3.3cm}
\includegraphics[width=3.2cm]{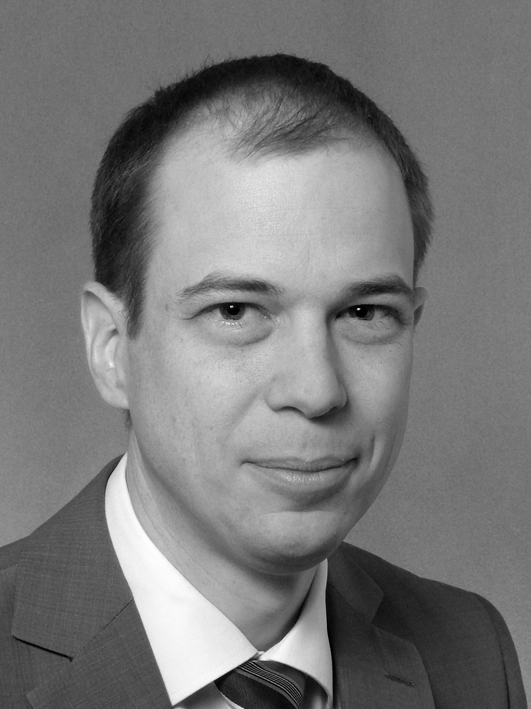}
\end{minipage}

\begin{abstract}

Quantum mechanical tunneling of atoms is increasingly found to play an important
role in many chemical transformations. Experimentally, atom-tunneling can be
indirectly detected by
temperature-independent rate constants at low temperature or by enhanced
kinetic isotope effects.
On the contrary, using computational investigations
the influence of tunneling on the reaction rates
can directly be monitored.
The tunnel effect, for example, changes reaction paths and
branching ratios, enables chemical reactions in an astrochemical environment
that would be impossible by thermal transition, and influences biochemical
processes.

\end{abstract}

\section{Introduction}

The tunnel effect is the quantum mechanical phenomenon that particles can
penetrate and pass areas in configuration space with a potential energy higher
than their total energy. While phenomenological descriptions appeared earlier,
the effect was discovered and understood in 1927 by Hund.\cite{hun27}
Subsequently, Gamow\cite{gam28} and Gurney \& Condon\cite{gur28} used the
tunnel effect independently of each other to explain the $\alpha$ decay of
atomic nuclei.  It was understood early on that the process of tunneling can
contribute to chemical reaction rates in addition or as an alternative to the
thermal barrier crossing.

\begin{figure}[ht!]
 \centering
 \includegraphics[width=8cm]{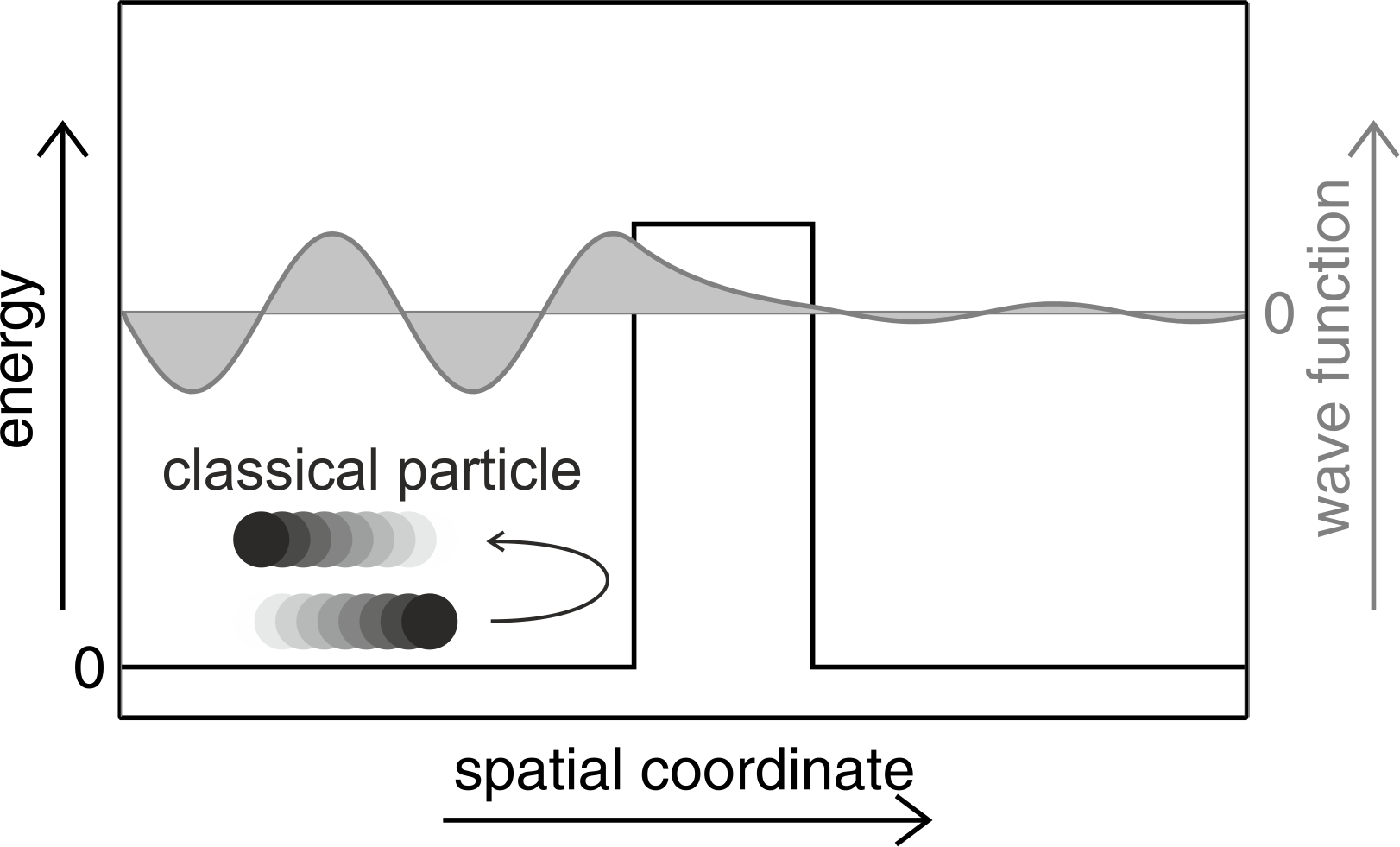}
 \caption{Wave function during the tunneling through a rectangular
   barrier. \label{fig:recbarr}}
\end{figure}

A number of quantum mechanical effects can be found in the movement of atoms
in chemical systems. Examples are the quantization of vibrational and
rotational energies, the zero point vibrational energy, and the tunnel effect.
The former two lead to line spectra and Fermi resonances, the latter two can
influence the rate constants of chemical reactions. Since atoms, like all
matter, have particle properties as well as wave properties, they have a
finite probability of appearance in regions in configurational space which
would be classically forbidden due to a potential energy which is higher than
the total energy of the atom, see \figref{fig:recbarr}. If such a region is
narrower than the extent of the particle wave, there is a finite probability
for the particle to appear at both sides, i.e. reactant and product sides. This
is the cause of the quantum mechanical tunnel effect. Thus, atoms can tunnel
through potential energy barriers.

The basic properties of the tunneling
probability can be easily analyzed by assuming a rectangular barrier of height
$E^\ddag$ above the energy of the particle, as depicted in
\figref{fig:recbarr}. The wave function within the barrier region is
proportional to $\exp(-x\sqrt{mE^\ddag})$.
The width of the barrier ($x$) enters linearly in the exponent, while
only the square roots of both the mass ($m$) and the barrier height
($E^\ddag$) enter. Similar
equations hold for differently shaped barriers. This simple example illustrates
the strong dependence of the tunneling rate on the barrier shape and somewhat
weaker dependence on the mass and the barrier height, in contrast to the
thermal rate described by the Arrhenius equation, which mainly depends on
  the barrier height.

There is the almost philosophical question how the particle can be present in
an area for which its energy is not sufficient. In principle, in the
classically forbidden region, the total energy is lower than the potential
energy of the particle, i.e., the kinetic energy would be negative. That is not
only counterintuitive but also physically meaningless. The paradox can be
resolved by looking at the relevant time- and length scales. Finding a
particle within the classically forbidden region with a known kinetic energy
would violate Heisenberg's uncertainty principle. In fact, the uncertainty in
these two quantities is always large enough that either the position or the
kinetic energy (via the momentum) escape the classically forbidden region.
 
\begin{figure}[ht!]
 \centering
 \includegraphics[width=8cm]{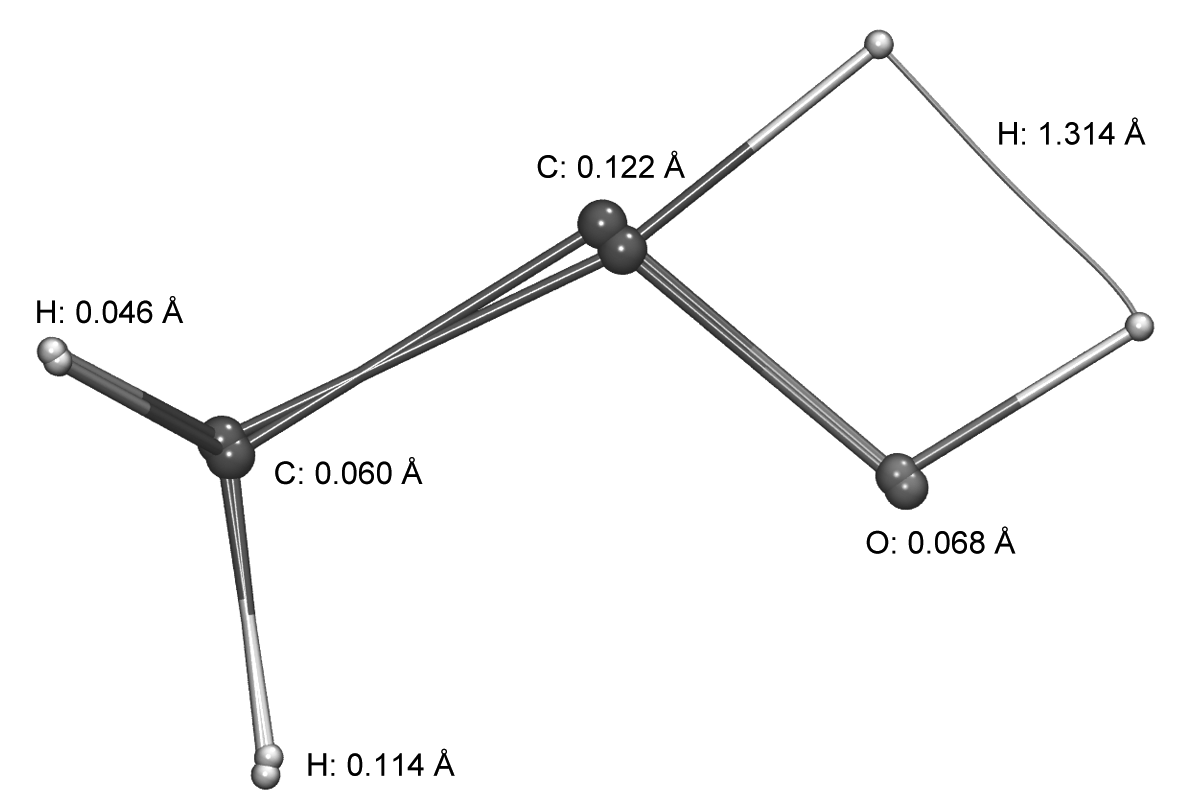}
 \caption{Displacement of atoms during the tunneling process from
   methylhydroxycarbene to acetaldehyde. The length of the tunnel path is
   indicated for each atom. While the motion is dominated by one hydrogen
   atom, it is obvious that the whole molecule contributes to the
   tunneling motion. Reproduced from reference \citenum{kae13}.
   \label{fig:tunnelpath}}
\end{figure}

At room temperature, atom tunneling is mostly relevant for hydrogen atoms due
to the mass-dependence of the tunnel effect. Usually many atoms of a molecule
move during a chemical reaction.  A typical reaction path (either
  classical or including tunneling) involves motions of several atoms other
than hydrogen, see \figref{fig:tunnelpath}.  Therfore, it is generally
impossible to assign a specific mass to a particular reaction. Even an
effective mass frequently changes during the course of a
transformation. Accordingly, many (or all) atoms of a molecule involved in a
specific reaction are tunneling which makes it generally difficult to
distinguish between hydrogen tunneling and heavy-atom tunneling, see
\figref{fig:tunnelpath}. There is no doubt, however, that the mass-dependence
of the tunneling rate leads to large kinetic isotope effects (KIEs). A KIE is
the ratio of the reaction rate of two isotopologues or isotopomers of a reaction, the rate of
the heavier isotopologue is divided by the rate of the lighter one. This
normally leads to KIEs larger than one even without tunneling. Values $<1$ are
called inverse isotope effects. Strong KIEs are the main experimental
indication for atom tunneling to happen. 
In the case of chemical reactions where one single atom -- mostly a hydrogen atom --
is transferred, e.g. in sigmatropic [1,5] H-shifts or hydrogen abstraction reactions, one can 
distinguish between a primary KIE and secondary KIEs:
The primary KIE is defined as the KIE arising from the substitution of the 
transferred atom by the heavier isotope.
Secondary KIEs are thus defined as KIEs arising from substitution of other atoms than
the transferred one by the heavier isotope.

\begin{figure}[ht!]
 \centering
 \includegraphics[width=6cm]{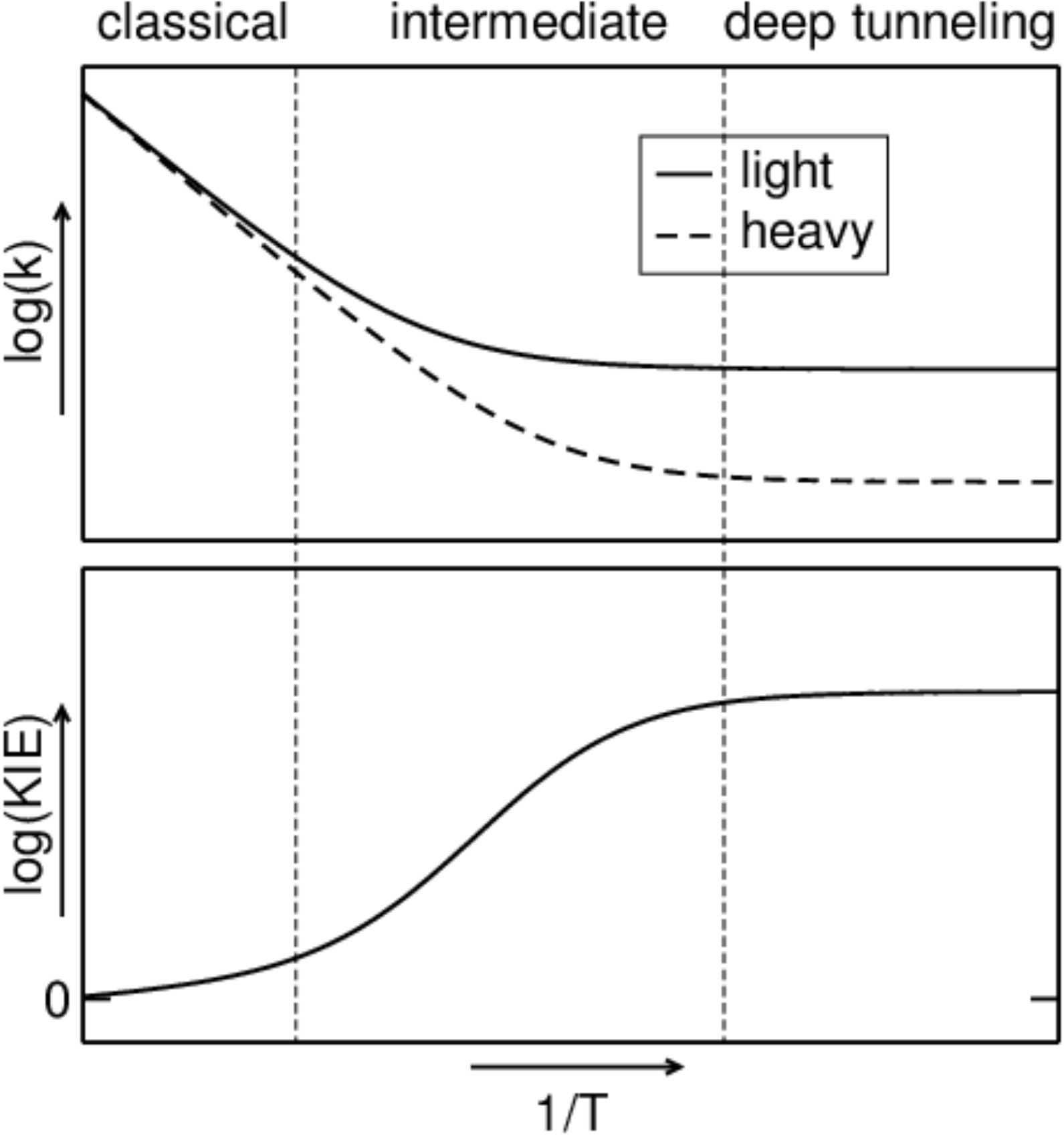}
 \caption{Upper part: Different temperature regimes of atom tunneling for two
   isotopologues shown in an Arrhenius plot: the logarithm of the rate
   constant is plotted against the inverse temperature.
   Lower part: The resulting logarithmic KIE as a function of inverse temperature.
   \label{fig:areas}}
\end{figure}

\Figref{fig:areas} shows the typical temperature-dependence of rate
constants and KIEs. At high temperature $k(T)$ follows the Arrhenius law, which results
in a linear Arrhenius plot. The intermediate temperature regime marks the
onset of tunneling and a curved Arrhenius plot. At low temperature, in the
deep tunneling regime, the reaction rate is
temperature-independent as tunneling occurs exclusively from the ground state. \Figref{fig:areas} actually shows the rate constant
for the unimolecular reactions H$_3$C--C--OH $\rightarrow$ H$_3$C--CHO (light)
and H$_3$C--C--OD $\rightarrow$ H$_3$C--CDO (heavy), details can be found
elsewhere.\cite{kae13} Tunneling is more efficient for the light  isotopologue
which causes a large KIE, especially at low temperature. 

Tunneling is most relevant at low temperature. At high enough
temperature, any reaction will be dominated by thermal transitions. The
  crossover depends on the specific reaction. For hydrogen transfer reactions
  it is often around room temperature. In intermediate temperature regimes,
the overall reaction rate is often found to be higher than the (extrapolated)
thermal rate and the low-temperature limit of the tunneling rate
combined. This phenomenon is referred to as temperature-assisted tunneling, or
vibrationally activated tunneling (VAT).\cite{dew85a}

The tunnel effect and its relevance for chemistry had been covered in many
reviews and even textbooks 
in the course of the majority of the last century.
The main reference
work is probably the textbook by Bell,\cite{bel80} other books and reviews
followed it.\cite{miy04,koh05,all09,koh03,nag06,bor16} Further review articles deal
with special aspects like atom-tunneling in enzymes\cite{lay14,joh15,var15} or
methods to calculate tunneling rates.\cite{fer06,pu06,nym14,kae14} Obviously,
quantum effects like tunneling also occur in many other areas than
chemistry. Electrons tunnel much more readily than atoms due to their lower
mass. This enables scanning tunneling microscopy, tunnel junctions, and tunnel
diodes. All these are outside the scope of this review, which summarizes the
development of the field of atom tunneling in the last decade with focus on
its consequences for chemistry.

\section{Methods to determine Atom Tunneling}

\begin{figure}[ht!]
 \centering
 \includegraphics[width=8cm]{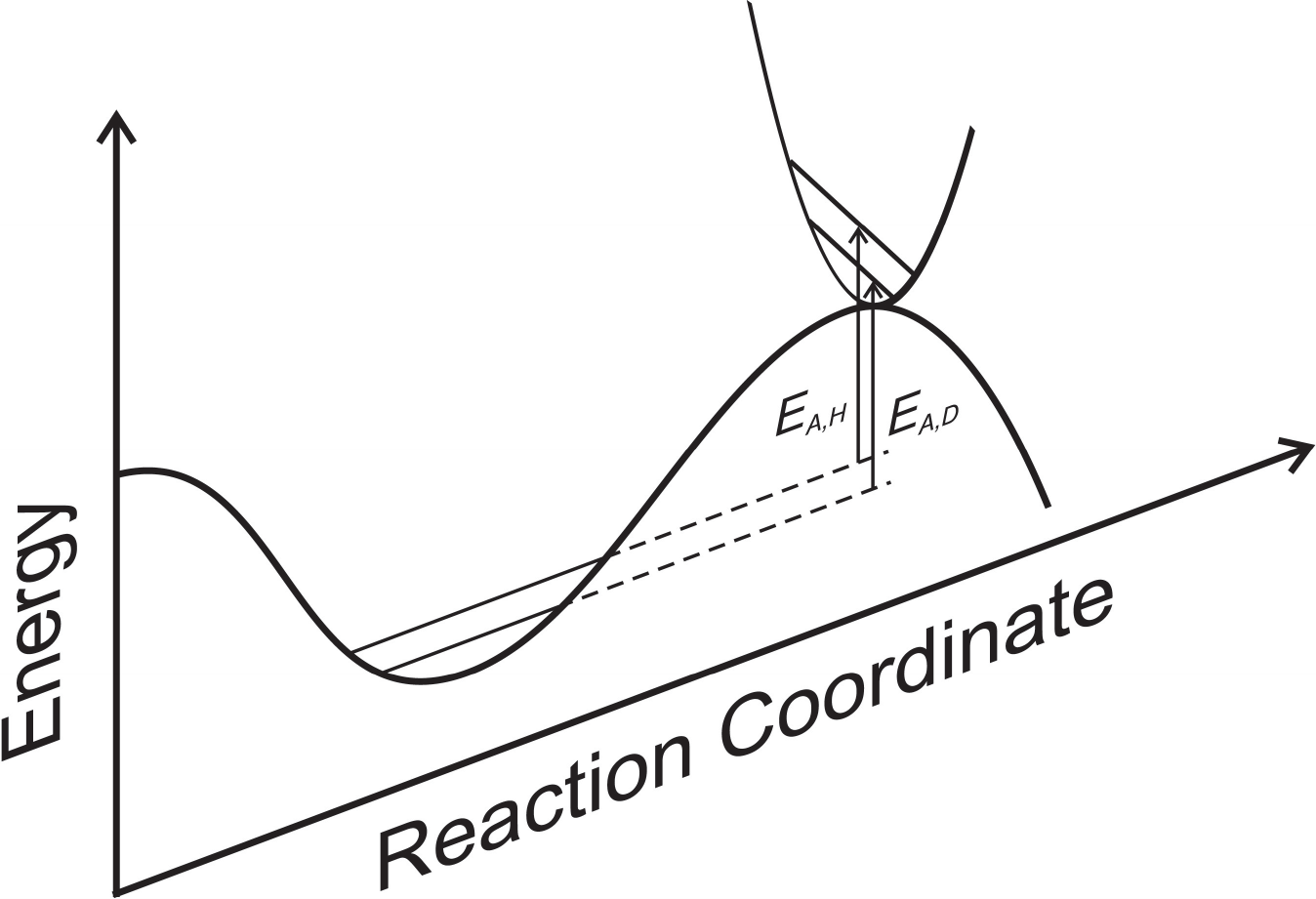}
 \caption{The effect of the mass on the zero point energy on a potential
   energy surface. Schematic drawing.
   \label{fig:pes_scheme}}
\end{figure}

Atom tunneling happens to some extent in any chemical reaction. At high
temperature, however, its contribution to the rate is negligible. An
experimental quantification of the tunneling effect is impossible, whereas in
simulations the effect can easily be switched on and off. The most important
technique to experimentally assess the importance of tunneling is to measure
KIEs, which are caused by tunneling and the zero point vibrational energy
(ZPE). The effect is illustrated in \figref{fig:pes_scheme}. While the
potential energy of the atomic movement is independent of the mass of the
atoms, the vibrational frequencies, and consequently the ZPE, are
mass-dependent. Higher masses lead to lower ZPEs in both the reactant
  state and the transition state. In unimolecular reactions, this change is
  generally larger in the reactant state than in the transition state,
  partially because there is one more vibrational mode in the former and the
  modes perpendicular to the reaction path are often stiffer in the reactant
  state. The difference in ZPE normally increases the vibrationally adiabatic
  barrier (i.e. potential energy plus ZPE) for the reaction of the heavier
  isotopologue, and thus reduces the reaction rate. Primary KIEs between
protium and deuterium of up to 6--7 at 300~K can be explained by the
difference in ZPE between H and D isotopologues, see
\figref{fig:pes_scheme}. If a higher KIE is found, this generally indicates
tunneling. Since specific values depend on the system under study, more
broadly applicable criteria were proposed. Fitting kinetic data with an
Arrhenius expression $k(T)=A\exp(-E_\text{A}/k_\text{B}T)$ results in a
pre-exponential factor $A$ and an activation energy $E_\text{A}$, which
results in a straight line in an Arrhenius plot as in \figref{fig:areas} in
the classical regime. Here, $k_\text{B}$ is Boltzmann's constant and $T$ is
the absolute temperature. Note, that the activation energy is not the height
of the potential energy barrier. It merely reflects the slope of the Arrhenius
curve. In deep tunneling, $E_\text{A}$ vanishes since the rate becomes
temperature-independent. The following approximate criteria have been proposed
for H/D-KIEs:\cite{kim92b} a difference in activation energy
$E_\text{A}(\text{D})-E_\text{A}(\text{H})>5.0$~kJ~mol$^{-1}$ and a ratio of
pre-exponential factors of $A(\text{H})/A(\text{D})<0.7$. Other criteria, like
Swain--Schaad exponents,\cite{swa58} despite frequent
use,\cite{cha89,koh03,kli10,koh10,mey11,ley12} were shown to not always be
suitable as indicators for tunneling.\cite{kim92b,fra02,pel08a} Techniques to
measure KIEs, especially in biochemical systems, are reviewed
elsewhere.\cite{koh10} All these can, of course, only serve as indications
since no reaction will proceed exclusively via atom tunneling or exclusively
without it.

In theoretical studies, atom tunneling can be switched on and off and its
effect on reaction rates can directly be monitored. Different techniques exist
to calculate reaction rates including atom tunneling. The simplest ones are
based on harmonic transition state theory\cite{eyr31} (HTST) and the multiplication of the
HTST-rate constant by a tunneling correction factor $\kappa$. It can be
obtained by assuming a specific functional form for the potential energy along
the reaction coordinate, for which $\kappa$ can be obtained analytically. The
Eckart barrier\cite{eck30} models some bimolecular reactions quite
realistically, simpler forms are a parabola\cite{bel35} or a simple
rectangular barrier. The latter is, despite its obvious shortcomings, still
sometimes used in astrochemical modeling.\cite{wat08,oba12,min13,con14a} These simple
approximations assume, however, that the tunneling process occurs along the
same reaction coordinate as the thermal reaction. They are sometimes referred
to as one-dimensional tunneling.\cite{pu06}

In reality, atom tunneling leads to corner cutting on the potential energy
surface.\cite{mar77} This is taken into account in multidimensional
  tunneling methods\cite{tru03,fer07} like the small-curvature
tunneling correction (SCT),\cite{sko81} a popular and successful method to
approximate tunneling rates, or in the large-curvature tunneling correction
(LCT)\cite{gar83,gar85} and more recent methods of a similar
basis.\cite{liu93b,alh01,mea10} The tunneling path is fully optimized in the
Feynman-path-based instanton
theory,\cite{lan67,lan69,mil75,col77,cal77,gil77,aff81,col88,han90,
  ben94,mes95,ric09,kry11,alt11,rom11,rom11b,kry14} sometimes also referred to
as harmonic quantum transition state theory (HQTST).\cite{and09} Ring-polymer
molecular dynamics,\cite{cra05,men14} or a special case thereof, centroid
molecular dynamics,\cite{cao94,vot96,pol98,ric09} and the related quantized
classical path method,\cite{hwa91,hwa93,maj07,azu11} the centroid density
method,\cite{gil87,vot89,vot93} and the reversible action-space work QTST
(RAW-QTST)\cite{mil97,mil98} are all based on Feynman's path integral
formulation.\cite{fey48} Quantum dynamics in the form of wave packet
dynamics,\cite{gar95} the multi-configuration time-dependent Hartree (MCTDH)
approach\cite{man92,mey90,pad08,ham09} and several other methods, where only a
few can be mentioned here,\cite{han96,che03} were used to simulate atom
tunneling as well. Computationally demanding methods like wave packet
  dynamics and MCTDH are typically applied on potential energy surfaces,
which need to be fitted in advance.\cite{mar11}

It is worth to mention an inconsistency in the semantics in the literature. KIEs
which can be explained by differences in the vibrational zero point energy
(ZPE) without tunneling are sometimes regarded as being explicable by \emph{semiclassical}
approaches. This terminology, probably introduced by Bell\cite{bel74,bel80}
and widely used in the biochemical literature, is rather unfortunate since the
physics community generally terms methods related to the WKB
(Wentzel--Kramers--Brillouin)\cite{ray12,jef24,wen26,kra26,bri26} as
semiclassically. The latter include SCT, LCT and the most common variant of
instanton theory and are perfectly able to describe tunneling phenomena.

\section{Impact of Tunneling on Different Fields of Chemistry}

Atom tunneling has been found in many different areas of chemistry. Here, we
use the most prominent examples to illustrate common concepts and facilitate
the interpretation of results. Since, e.g., astrochemistry mostly happens at
cryogenic temperature while only ambient temperature is relevant for
biochemistry, the implications of tunneling on these fields is fundamentally
different. We aim at depicting these differences as well as common concepts.

\subsection{Biochemistry}

The majority of enzymatic reactions involve a hydrogen transfer step: hydride,
hydrogen radical or proton transfer. At room temperature most hydrogen
transfer reactions are influenced by tunneling, at least to some
degree. Therfore, it is obvious that tunneling is a vital component in many
biological processes. Experimental evidence has been found mainly through
large H/D-KIEs.\cite{cha89,gra89,koh05,all09,kli10} Values above the range
that can be explained by differences in the zero point energy alone have been
found in dozens of enzymes, the most prominent being
lipoxygenases,\cite{gli95,mey11} taurine/$\alpha$-ketoglutarate dioxygenase
(TauD),\cite{pri03} and aromatic amine dehydrogenase (AADH).\cite{mas06} A recent
more extensive list is given by Klinman.\cite{kli10} The temperature window
accessible by biochemistry is narrow, but still H/D-KIEs $>500$ were
observed.\cite{hu14} The tunnel effect can directly be quantified in theoretical
investigations. Probably the most promising\cite{sen09a} method to simulate
enzymatic processes is the QM/MM approach.\cite{war72,war76} It has been used
to confirm several experimentally found KIEs{\cite{gao02,tru04,ran10,var15}} and shed light on the
mechanistic implications of tunneling. Cluster models and the investigation of
surrogate systems as well as the reaction specific fit of force fields like
the EVB approach, were also applied on a broad scale.\cite{ols04a,vil01,liu07}
Overall, it is clear that atom tunneling does occur in biological systems,
albeit not causing the main catalytic effect.\cite{wil10}

KIEs can in many cases be estimated quite accurately by
theory,\cite{vil01,liu07,maj09,rom12,aba13,aba14,zen15} while absolute
reaction rates are often in poor agreement between theory and experiment. In
many cases, this can be explained by different quantities being
compared. Computational simulations of enzymatic processes focus on the
chemical step of an enzymatic process.  This should not be compared to, e.g.,
the experimental turnover rate $k_\text{cat}$, which often represents the
product release.  The whole process contains many more elementary steps that
can become rate limiting, e.g. diffusion, substrate binding, conformational
changes, or product release.\cite{koh15} They all have their different
kinetics and most of them are isotopically insensitive. Thus, the apparent KIE
on either $k_\text{cat}$ or the catalytic efficiency $k_\text{cat}/K_\text{m}$
differs from the intrinsic KIE of the isotope-sensitive step. The intrinsic
KIE can be estimated by using commitments to forward and reverse
reactions.\cite{her82,coo91,koh03,koh05,koh10} Careful comparison between
theory and experiment results in excellent agreements. For example in
dihydrofolate reductase (DHFR), experimental intrinsic H/D
KIEs\cite{sik04,sen11} and values calculated on high level\cite{aga02,fer03a}
both result in a KIE of 3.5 $\pm$ 0.1, independently of the temperature and
almost independently of the pH.{\cite{pu05}} Apparent KIEs, measured with
different techniques result in KIEs $<3.0$,\cite{lov12} strongly indicating
that other steps than the chemical transformation mask the KIE.\cite{koh15}

A particular challenge in the description of biochemical proton
transfer is that the proton transfer is often coupled to an electron
transfer (PCET).\cite{lay14,ham15a} Both, proton and electron move
according to quantum mechanics. In many cases the electron movement is
much faster than the proton movement, though.

The temperature-dependence of H/D-KIEs in enzymes has been measured in many
cases. It turned out that the intrinsic KIE of almost any natural enzyme is
rather temperature-independent,\cite{kli10,nag09,koh15} at least over the
limited temperature range available to biochemical reactions. In many mutants,
a temperature-dependence, also of intrinsic KIEs, was found. A
temperature-independent KIE could be explained by deep tunneling,
i.e. tunneling out of a single quantum state rather than from a thermal
  ensemble, see \figref{fig:areas}. For enzymes, this explanation does
  not apply, however, since the rate constant itself still strongly depends
on the temperature. The observations have been rationalized by a pre-tunneling
state\cite{lim10} with an energy above that of the reactant state but below
the barrier. Tunneling occurs from that pre-tunneling state rather than from
the bottom of the reactant well. A high-lying pre-tunneling state causes
tunneling from that state for both H and D at ambient temperature, which
explains moderate, temperature-independent KIEs and a temperature-dependent
rate constant.

There is a longstanding and partially heated debate whether or how protein
vibrations, which couple to the hydrogen transfer reaction coordinate, enhance
enzymatic H-transfer reactions. The proposal is\cite{joh15} that a slow global
reorganization of the enzyme structure forms a tunneling-ready configuration,
possibly by proper alignment of quantum states which allows efficient
  resonant tunneling. Orthogonal to that slow
reorganization, the system moves fast along the hydrogen transfer coordinate
in the tunneling-ready configuration.  Promoting vibrations were
proposed,\cite{ant97,ant02} which lead to donor-acceptor compression on the
timescale of barrier crossing and increase the tunneling
probability. To be effective, such promoting vibrations must be very fast,
comparable to C--H stretching frequencies.\cite{joh07} This may be possible if such
vibrations are highly localized.\cite{hat07,joh07} However, they were used to
explain the effects of remote mutations.\cite{ant02,pud09} At present, there
are only indirect indications, but no direct experimental evidence, that such a
vibronic model can explain the temperature-dependence of KIEs.\cite{joh15} No
detectable dynamic coupling of protein motions to the hydrogen transfer step
was found experimentally in DHFR.\cite{lov12} Barrier compression was claimed
to favor quantum effects in the catalysis,\cite{hay10} however, other
researchers found that it enhances the reaction rate mostly by lowering the
barrier which, in fact, reduces the amount of tunneling compared to the
thermal rate.\cite{kam10,kam10a} Comparison of the turnover rate in a
wild-type enzyme with its heavily deuterated counterpart (``heavy enzyme'')
showed that no specific protein motions are responsible for enhancing tunneling.\cite{luk13} Part of the current dispute is probably
semantics: while a vibrational model can be set up by using exclusively
equilibrium dynamics,\cite{koh15} non-equilibrium (non-statistical) dynamics
was sometimes employed which raised criticism.\cite{kam10,kam10a,vil01}
Pressure-dependence of KIEs was used to argue in favor of the vibronic
model,\cite{hay07,hay12} but was shown not to provide evidence that promoting
vibrations enhance the catalytic effect.\cite{kam10,kam10a,kam10b}

\subsection{Barrier Width}

In chemical reactions, tunneling is most efficient if light atoms move just a
short distance during the rate determining step of a reaction as the barrier
width determines the probability of tunneling: The narrower a barrier is the
the more likely is tunneling.\cite{kar15,kae13,cam05,joh02} It was shown
decades ago that in the reaction of 2,4,6-tri-\emph{tert}-butylphenyl radical
to 3,5-di-\emph{tert}-butylneophyl, see \schemeref{fig:tbu}, the H atom has to
travel just a small distance of 1.34 \AA \ in the transformation from the
reactant to the product.\cite{bru76} This leads to a kinetic isotope effect of
13,000 at $-150$\textcelsius{} and still 80 at $-30$\textcelsius.

\begin{scheme}[ht!]
 \centering
 \includegraphics[width=8cm]{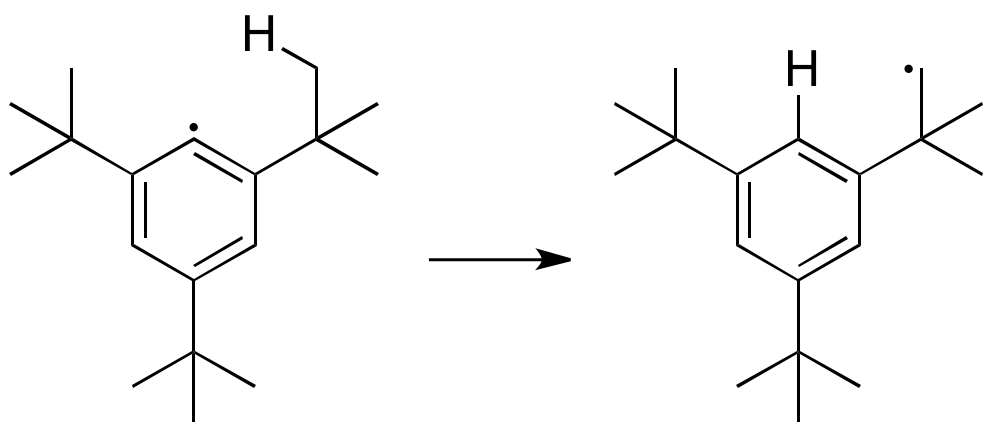}
 \caption{Reaction of 2,4,6-tri-\emph{tert}-butylphenyl to 3,5-di-\emph{tert}-butylneophyl.
   \label{fig:tbu}}
\end{scheme}

Carbenes are generally known to be highly unstable molecules.  For a few
simple ones, namely hydroxycarbene (H--$\ddot{\text{C}}$--OH),
methylhydroxycarbene (Me--$\ddot{\text{C}}$--OH), and phenylhydroxycarbene
(Ph--$\ddot{\text{C}}$--OH) among others, it was shown that they are unstable
even at cryogenic temperature.\cite{sch08,buc08,ger10a,zue03,wan15} A [1,2]
H-shift to formaldehyde, acetaldehyde, or benzaldehyde, respectively, is
enabled by tunneling because of the small distance the corresponding hydrogen
atom has to surmount.  By contrast, dihydroxycarbene
(HO--$\ddot{\text{C}}$--OH) does not react to the respective product, formic
acid, probably because of the strong $\pi$ donation of the oxygen atoms
reducing the electron deficiency of the carbene.\cite{sch08a}
Methylhydroxycarbene exhibits two different reaction pathways:\cite{sch11a}
one results in vinyl alcohol, the other in acetaldehyde.  The energy barrier
of the former reaction is lower, while the one of the latter is narrower.  At
high temperatures, the classical thermal reaction causes vinyl alcohol to be
formed preferentially, while at low temperature, tunneling happens through the
thinner barrier and acetaldehyde is formed.  This nicely shows the impact of
the shape of potential energy barriers, in particular the length of reaction
paths on tunneling. The concept that the barrier width determines the
tunneling probability lead to Schreiner's formulation of \emph{tunneling
  control} of chemical reactions:\cite{ley12,ley13} whereas the concepts of
thermodynamic control (the lowest-energy products will be formed in the
long-time limit) and kinetic control (the reaction with the lowest barrier
happens first) are widespread, tunneling determines the selectivity at low
temperatures.  Further examples of tunneling control are, e.g., in the ring
expansion of noradamantylcarbenes:\cite{mos04,koz13} different substituents
change the reactivity at cryogenic temperatures and suppress tunneling or lead
to different products.\cite{koz13,koz14b}

The question if rather the [1,2] H-shift to the alkene or 
to the corresponding aldehyde takes place in different
hydroxymethylcarbene-analogs was answered computationally, concluding that
the tunneling path length is the decisive quantity.\cite{kae13} Using
cyclopropylcarbene or 1-methylcyclobutylhalocarbenes the possibility of
[1,2] H-shifts is suppressed. Instead, ring insertion reactions take
place, facilitated by carbon tunneling, see
\schemeref{fig:carbontunnel}.\cite{ger11,zue03,ley11}

\begin{scheme}[ht!]
  \centering
  \includegraphics[width=5cm]{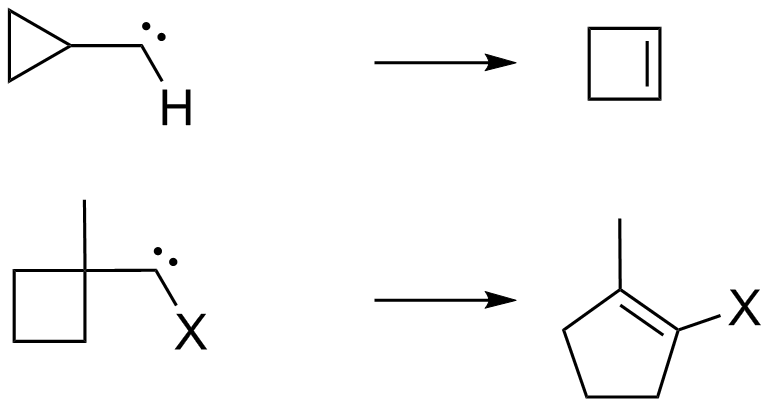}
  \caption{Ring expansion due to carbon tunneling observed in
    carbenes.\cite{ger11,zue03,ley11} X=Cl,F
    \label{fig:carbontunnel}}
\end{scheme}

\subsection{Organic Chemistry}

In organic chemistry hydrogen atoms are frequently found to tunnel, even at
room temperature.  In the case of reactions of closed-shell molecules with
radicals, hydrogen is abstracted if the emerging radical is more stable than
the previous one.  Various reactions of small organic molecules with hydrogen
atoms,\cite{sha14,wan14,oue15,cao14} hydroperoxyl radicals,\cite{men14}
chlorine atoms,\cite{cho14,bai15,ng15,li14} or the activation of
H$_2$\cite{hen15} have been studied experimentally as well as computationally.
Even at room temperature these reactions are influenced by tunneling, raising
the reaction rate constants. For Claisen rearrangements it was necessary to
computationally include a model of tunneling through a parabolic barrier to explain the
experimental $^{13}$C KIEs\cite{mey99,kup92,kup93} and in a Swern oxidation,
multidimensional tunneling had to be included in computations to reproduce the experimental
results.\cite{gia10}

Several studies of the tautomerization of small or medium-sized molecules show
the impact of tunneling on proton shifts.  Tunneling decay of particular
conformers of glycine,\cite{baz12a,baz12b} alanine,\cite{baz13}
cytosine,\cite{rev12} and other small molecules with relevance to
biology\cite{mai12,ger14} was observed as well as the tunneling-accelerated
tautomerization of tetrazole acetic acid.\cite{and14} 
Hydrogen peroxide is chiral if the rotation around the O--O bond is restricted
at low temperature. The stereo-mutation of
one enantiomer to the other one was investigated by
six-dimensional quantum dynamics\cite{feh07} and shown to proceed efficiently
by tunneling. Rotation of the OH-group in phenol was shown to happen via
tunneling by FTIR spectroscopy, while ortho- or meta deuteration suppresses
the tunneling motion at low temperature.\cite{alb13a} The \emph{cis-trans}
isomerization reactions of carboxylic acids was studied
extensively.\cite{pet97a,mar07d,ami10,lop10,baz12,tsu15,ger15,sch15} These
reactions are mostly carried out at cryogenic temperature in noble gas or
N$_2$ matrix environments.  It was shown that such an environment can
influence the tunneling rates of the isomerization reactions
significantly. This might be taken as a caveat when comparing to, e.g., quantum
chemical gas phase calculations.\cite{pet02,mac04a,mac05,dom09,lop10,baz12}

In bigger, bio-organic molecules, tunneling supports radical reactions like in the
autoxidation of tetraline.\cite{muc15a}
The regeneration of vitamin E (tocopherol) by ubiquinol was shown to be
accelerated more than 4000 times by tunneling.\cite{ina13}
Furthermore, high experimental KIEs denote that tocopherol-mediated peroxidation 
of fatty acids and 7-dehydrocholesterol might be promoted/supported by hydrogen tunneling.\cite{lam14,muc15}

Tunneling in different hydrogen bond networks has been investigated. 
One of the most interesting findings is the simultaneous proton tunneling in 
solid p-\emph{tert}-butyl calix[4]arene at low temperature.\cite{bro99}
Using NMR relaxometry it was possible to investigate the phonon-assisted tunneling
in the quadruple synchronous  proton transfer of calix[4]arenes, see \schemeref{fig:calix}.\cite{ued12,ued13} The
coupling of hydrogen bond dynamics to  large-amplitude motion in the vicinity
was studied by NMR.\cite{loz12}
  \begin{scheme}[ht!]
 \centering
 \includegraphics[width=7cm]{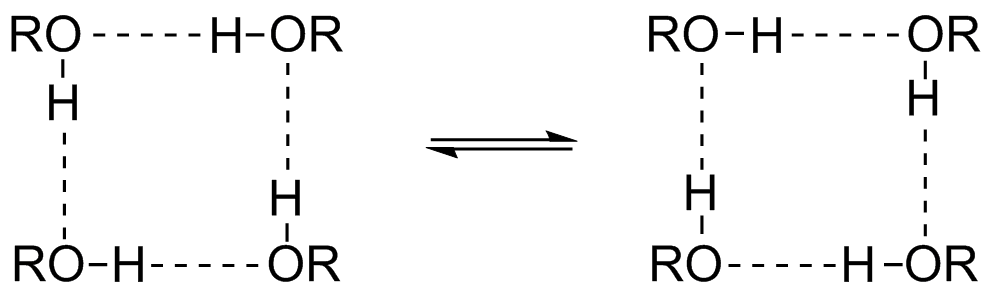}
 \caption{Quadruple proton transfer in calix[4] arenes.\cite{bro99,ued12,ued13}
     \label{fig:calix}}
\end{scheme}
Tunneling splittings caused by double proton transfer in various porphycenes
were observed.\cite{sme14,hom14,men15} In these cases the initial and final
state are equivalent: resonant tunneling takes place, which is particularly
fast. The flux of electronic and nuclear densities in a resonant tunneling
pericyclic reaction shows that only a rather small fraction of particles
actually has to move to accommodate such reactions.\cite{bre11} Rotation of
hydrogen-bonded water molecules was also shown to be facilitated by
tunneling.\cite{sch14,alv14} The fluctuation between hydride and dihydrogen
ligands of Fe$^\text{II}$ was demonstrated to be dominated by tunneling using
quasi-elastic neutron scattering and computational investigations.\cite{dos11}

\begin{figure}[ht!]
  \centering
  \includegraphics[width=8cm]{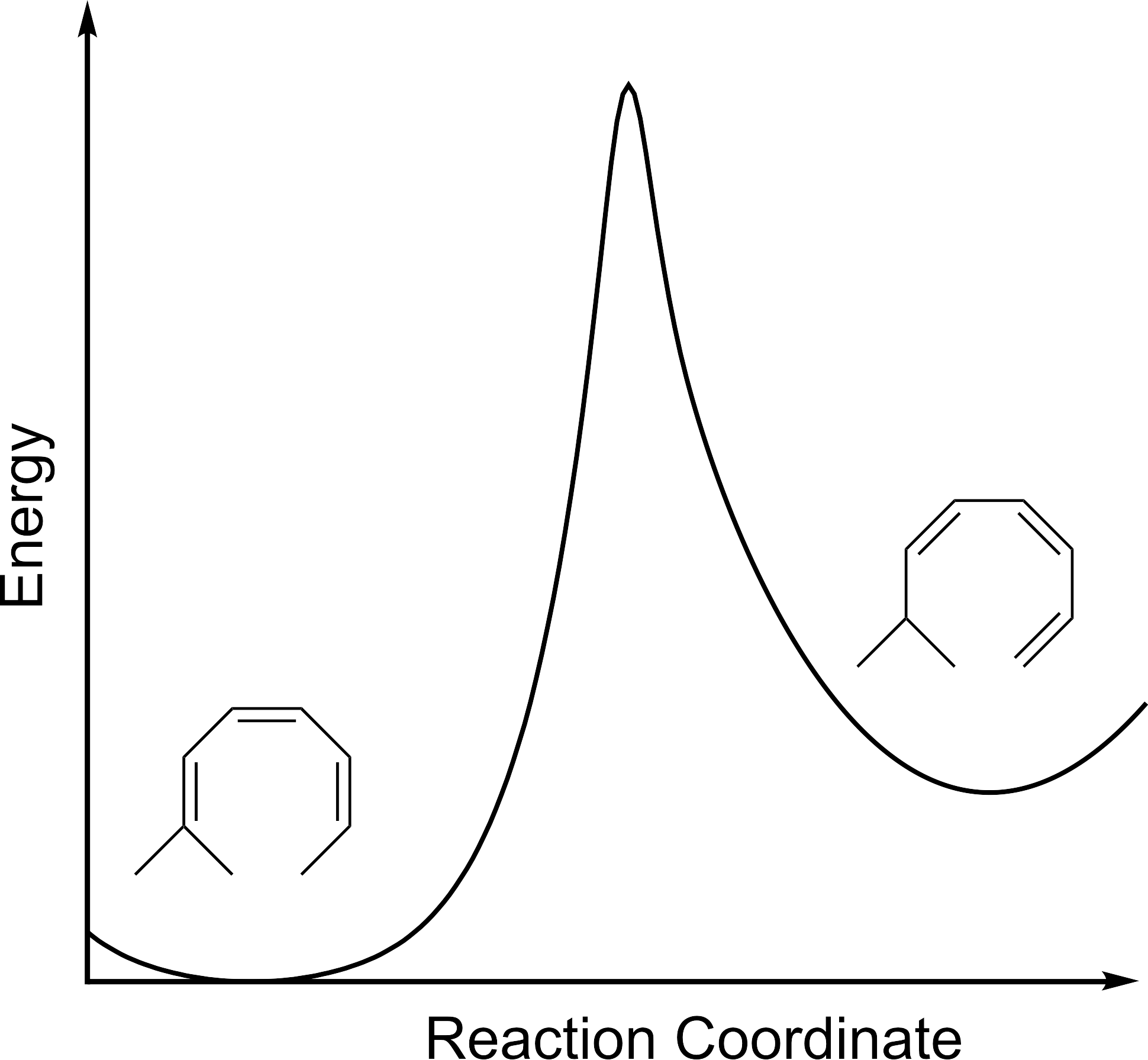}
  \caption{Schematic reaction profile of the [1,7] sigmatropic hydrogen shift
    of 7-methylocta-1,3(Z),5(Z)-triene to
    2-methylocta-2,4(Z),6(Z)-triene. The sharply peaked barrier
    facilitates tunneling.
    \label{fig:17shift}}
\end{figure}

In sigmatropic rearrangements, tunneling was observed in many cases.
Suprafacial [1,5] sigmatropic rearrangements were studied
exhaustively, using derivatives of
1,3(Z)-pentadiene.\cite{dor86,liu93a} Although, initially it was
unclear if tunneling plays a crucial role,\cite{doe06,doe07} various
studies meanwhile confirm involvement of tunneling.
\cite{liu93a,van07,she07,pel08b,zim10,kry12,kry14} An antarafacial [1,7]
sigmatropic hydrogen shift can happen in 1,3(Z),5(Z)-heptatrienes, see \figref{fig:17shift}.
Using 7-methylocta-1,3(Z),5(Z)-triene as model system, it was shown
that the isomerization from provitamin D 
to vitamin D is accelerated
by tunneling.\cite{bal88,mou07,mea12,kry14} For the antarafacial [1,7]
sigmatropic hydrogen shift the hydrogen atom just moves a small
distance in the high-energy region of the reaction, although the
potential energy minima are quite far away because of reorganization
of the carbon framework.  

Despite the large number of studies, further investigation is necessary to
elucidate the impact of multidimensional tunneling on the different
variants of sigmatropic rearrangements or other pericyclic reactions.

\subsection{Catalysis}

In organometallic chemistry and homogenous catalysis, some reactions show high
KIEs, indicating tunneling to be important for the rate determining step.
\cite{sla10,fuk10} One of the most impressive cases might be a hydrogen
exchange reaction in a titanium complex: here, at 200~K the exchange of a
$\beta$ H-atom is suppressed when deuterium is used.\cite{dun13} This is
equivalent to a KIE of $>16,000$.  The homolytic
cleavage of a C--H bond enforced by an osmium centered radical at
25\textcelsius{} has a KIE larger than 16.\cite{lew14} Also, the protonolyses
of palladium and platinum complexes and the reductive elimination of methane
from a gold complex have shown to have a significant tunneling
contribution.\cite{ber08,ber10,sco11,nij14}
Tunneling is even more pronounced in different reactions of bio-mimetic model 
complexes involving iron in high oxidation states:
these exhibit large H/D KIEs, like in the case of a C--H hydroxylation reaction with an 
oxoiron(IV) porphyrin radical cation, showing a H/D KIE of 360  at $-30$\textcelsius{}.
\cite{pan08,con14,man15} 

At +23\textcelsius{}, a KIE of still 28 has been
reported for the same reaction.  High KIEs in oxoiron(IV) complexes
require a two-state reactivity model for
explanation.\cite{sch00b,sha07,kli09,kwo15} Although the ground state of the
reactant is a triplet, the low lying quintet state plays a significant role
since there the reaction barrier is significantly lower.  A C--H vibration
lowers the quintet state below the triplet state and a spin crossover during the
reaction was proposed.\cite{man15} In this way, a C--H bond length dependence
in reactivity can be explained.\cite{kli09} Analogously to these reactions
of the oxoiron compounds, an oxoruthenium(IV) complex featuring comparable
structural motifs has shown to display a KIE of 49 for the hydrogen
abstraction reaction of dihydroanthracene.\cite{koj11,dhu15}


Even though this review mainly focuses on molecular systems, we will now
briefly discuss atom tunneling on surfaces. Hydrogen atoms were frequently
observed to tunnel in surface processes.{\cite{lau85,ger13,hak10}} It was shown that the
motion of hydrogen on Cu(001),\cite{kua01,lau00,sun04,sun05}
Pd/Cu(111),\cite{kyr14,fir14} Ru(0001),\cite{mci13} W(110),\cite{dif80} and
Ni(100)\cite{lee92,wan09,sul12} surfaces is enhanced by tunneling at low temperature.
Even the motion of CO on a Cu(111) surface
was shown to occur below 6~K using scanning tunneling microscopy (STM), with a
temperature-independent hopping rate.\cite{hei02}

In the field of heterogenous catalysis, the CO oxidation and the dissociative
H$_2$O desorption as well as the OH dissociation on various metal (111)
surfaces were shown to be affected by tunneling.\cite{ger10a,ger12,ger07} The
dissociation of CH$_4$ on Pt(111) and Ni(111) was found to involve thermally
assisted tunneling.\cite{har91,jac13} Further, the dissociation and
recombination rates of H$_2$ on Ni(100) and the NH formation on a Ru(0001)
surfaces and the following successive H-addition reactions -- important in the
process of NH$_3$ formation -- are found to be accelerated by tunneling
especially at lower temperatures.\cite{wan13a,wal13,tau06,tau05,tau06a}
Even oxygen tunneling was
observed in the dissociative adsorption on an Ag(111)\cite{kun14}
and on Pt(111) surfaces.\cite{shi13}.
Heterogenous catalysis, however, often employs high temperatures where
tunneling is less important.

\subsection{Heavy-Atom Tunneling}

All atoms in a molecule are generally involved to some extent in the tunneling
motion, see \figref{fig:tunnelpath}. Thus, heavy-atom tunneling happens in the
reactions mentioned previously as well but usually plays a minor role.
Nevertheless, for a few well known textbook reactions, clear-cut heavy-atom
tunneling was shown to be involved.
While the increase in absolute reaction rate constants 
due to heavy atom tunneling is often rather small,
it leads to heavy-atom KIEs like $^{12}$C/$^{13}$C being
detectable by, for example, mass spectrometry.

In the Bergmann cyclization, carbon atom
tunneling accelerates the reaction rate by 38--40\% at 30\textcelsius{} as
obtained by DFT and CASSCF calculations.\cite{gre13a} In the Roush
allylboration of \emph{p}-anisaldehyde, multidimensional tunneling involving
heavy atoms is necessary to explain the experimental $^{12}$C/$^{13}$C KIEs at
$-78$\textcelsius{},\cite{vet12} a temperature commonly used in organic
synthesis. 
At this temperature the reaction is accelerated by a factor of 1.36 by heavy atom tunneling,
which can not be observed directly by experiments.\cite{vet12}
Thus, the $^{12}$C/$^{13}$C KIEs are a suitable probe to investigate the
agreement of experimental and theoretical methods.
Oxygen tunneling is found in the ring-opening reaction of cyclic O$_3$ to its
usual (open) form.\cite{che11} This rearrangement is observed even below
150~K.  In this temperature regime, the reaction rate constants are calculated
to be almost constant with a $^{16}$O/$^{18}$O-KIE of up to 10 for the
reaction of $^{18}$O$_3$.

Even though for these reactions tunneling of second-row elements perceptibly
contributes to the reaction, it generally plays a limited role for most
chemical reactions like catalysis or synthesis. Most chemical reactions
are performed at relatively high temperature and the conformational change during
the rate limiting step likely involves a significant motion of carbon, oxygen
or other heavy atoms.  However, at lower temperature where the
kinetic energy of the nuclei is too low to overcome the potential energy
barrier, heavy atom tunneling gets more important.  It is worth to study
chemical reactivity in the deep temperature regime to gain insight into the
elementary processes of kinetics, stability of molecules, and the nature of
atom tunneling.  For instance, heavy atom tunneling can spoil chemical
stability even close to 0~K as molecules expected to be stable classically
will decay by tunneling.  Examples for this are the rearrangement of
tetrahedryl-tetrahedrane or the decomposition of a hyper-coordinated
carbocation.\cite{koz14,koz15a}

A noteworthy case is the carbon tunneling in the automerization of the
antiaromatic cyclobutadiene and its derivates which probably is among
the first evidences for heavy atom tunneling in
chemistry.\cite{car83,dew84,hua84,car88,lef90} Tunneling in other anti-
or non-aromatic systems like the automerization reactions of pentalenes
and heptalenes or the isomerization of cyclopropenyl anions and the
impact of substituents was found more recently.\cite{koz14a,koz15}

Besides the reactions of the carbenes introduced above, other
automerizations, ring opening or closing reactions, and rearrangement
reactions -- in particular of strained molecules -- are also
influenced by heavy atom tunneling at low
temperature.\cite{buc79,spo89,hen12,bar13,inu13,ert14,kar14,kar15,koz15}
Reactions of strained organic molecules often involve unusually strong heavy atom
tunneling: a high activation energy due to the breaking of a C--C bond
combined with little movement of the involved atoms lead to narrow
barriers, which enhance the probability for tunneling. 
For instance, in the Cope rearrangement of semibullvalene\cite{zha10} or 
for the ring opening reaction of cyclopropylcarbinyl
radical\cite{dat08,zha09,mei11,gon10,zha11} heavy atom tunneling is prominent.
In the former case, $^{12}$C/$^{13}$C-KIEs of more than 5 at 40~K are
predicted.\cite{zha10} In the latter one, the reaction rate is nearly constant
below 20~K.\cite{mei11}

\subsection{Astrochemistry}

Astrochemistry describes the formation, distribution, and destruction of
chemical substances in space.  Noteworthy features, when considering
reactions and reaction rates in the interstellar medium, are the low
particle density and the strong radiation fields and generally the low
temperature.  Although more than 170 molecular species were detected
so far (not including isomers and isotopologues), all of them except fullerenes
are smaller than 14 atoms.
In diffuse clouds temperatures are
around 100~K and can be as low as just a few Kelvin  in dark clouds. 
\cite{dis13}
Thus, chemical reactions only occur
if they are barrierless, for instance induced by photons or cosmic rays, or via
tunneling. 

The de Broglie wavelength of particles increases with
decreasing momentum and, thus, with decreasing temperature.
Consequently, at the temperatures predominant in interstellar medium, atom tunneling
has to be considered for nearly all reactions featuring a potential
energy barrier, especially when hydrogen is involved. 

Many bimolecular reactions exhibit a pre-reactive minimum, a van-der-Waals
complex, in the entrance channel before the barrier. Such complexes increase
the attempt frequency for the reaction. Their lifetime increases with
decreasing temperature.  In combination with atom tunneling  the increasing
attempt frequency can even lead to an increase of the rate constant at
decreasing temperature.  Experimental evidence for this counterintuitive
effect exists in a few cases, for example in the gas-phase reactions
\begin{center}
H$_2$ + NH$_3^{+}\rightarrow$ NH$_4^{+}$ + H \cite{her91a}
\end{center}
 and 
\begin{center}
  CH$_3$OH + OH $\rightarrow$ CH$_3$O + H$_2$O.\cite{sha13}
\end{center}
In a full quantum mechanical picture, tunneling occurs from bound metastable
states.\cite{wei03}

Chemistry in space happens frequently on surfaces of dust
grains.\cite{tie82,boo15} These often consist of silicates or
carbonaceous compounds and are usually coated by frozen CO, water, methane and
other small molecules.\cite{tie82}

For example,
the surface reaction of hydrogen addition to CO, a key step in the formation of methanol
in space, is governed by tunneling, leading to strong H/D-KIEs.\cite{hir02,hid09,and11} 
Also for the hydrogenation of formaldehyde leading to methoxy radical,
tunneling is important.\cite{gou11b}

Many small molecules are found to be heavily deuterated in
space.\cite{par02,par04,cas12}
For some of them, like methanol\cite{gou11a} and formaldehyde,\cite{gou11b}
this can be explained by tunneling: 
the lighter protium can be abstracted by a hydrogen atom to form H$_2$
while deuterium remains bound to the COH$_x$ fragment. Subsequent
barrier-less recombination with another protium or deuterium atom leads
to deuterium enrichment, see \schemeref{fig:co}.

\begin{scheme}[htb!]
  \centering
  \includegraphics[width=12cm]{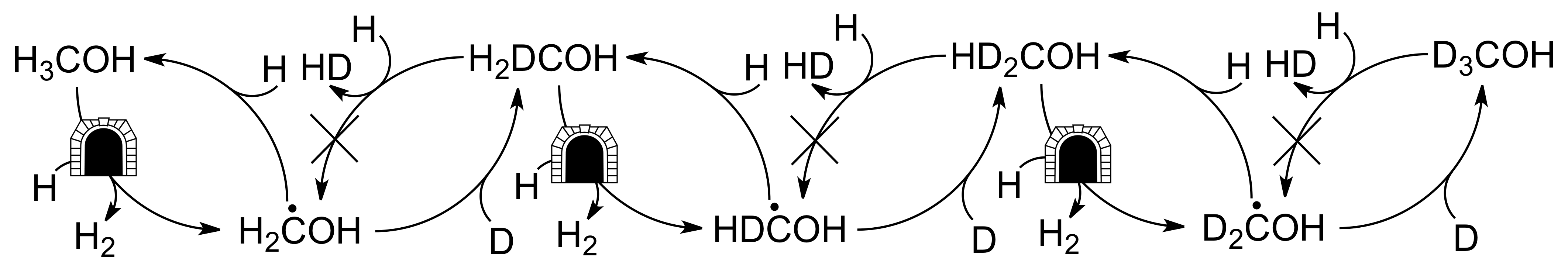}
  \caption{Reaction network for the deuteration of methanol. The abstraction of
    protium is facilitated via a tunneling mechanism, the abstraction of
    deuterium not.
    \label{fig:co}}
\end{scheme}
 
Although tunneling also takes place in reactions involving only non-hydrogen
atoms, like O + CO $\rightarrow$ CO$_2$, it was shown that in this case tunneling sets in
at a too low temperature to be of astrochemical importance.\cite{gou10a}

One model to simulate H$_2$ formation on carbonaceous dust grains, the
hydrogenation of benzene, was studied by quantum chemistry.\cite{gou10} Here,
tunneling contributes to the reaction rate of the first hydrogen chemisorption
while the addition of the second hydrogen atom is
barrier-less.\cite{gou10,gou11,ham15} Amorphous solid water is among the most
common surfaces in the interstellar medium as most dust grains are covered by
water.  It was found experimentally that water
can be formed from H$_2$ and OH on water surfaces even at 10~K,\cite{oba12}
even though the gas phase reaction exhibits a barrier of 17.5 kJ mol$^{-1}$.
For further reactions on water surfaces enhanced by atom tunneling, we refer
to reference \citenum{ham13}.

\section{Conclusions}

Quantum mechanical tunneling of atoms, despite it being known for almost 90
years, still provides challenging and surprising results as it influences
chemical reactions. While it is dominant at low temperature and for reactions 
involving atoms with low masses, hydrogen transfer reactions are 
often accelerated at room temperature and
above by the tunnel effect. Tunneling causes strong KIEs, making it detectable by
experiment. Simulations, on the other hand, are able to directly monitor the
tunneling process and can quantify its influence on the reaction rate
constant. Different computational techniques, from simple one-dimensional
corrections to classical TST over semiclassical approaches to full quantum
dynamics, are available. Symmetric reactions, where the reactant and product
are chemically indistinguishable, show resonant tunneling. It causes a
splitting of the vibrational energy levels. In contrast to that, thermal
reaction rates are typically influenced by non-resonant tunneling. Atom
tunneling was found in organic chemistry, inorganic chemistry, surface
science, astrochemistry and biochemistry. Astrochemistry is typically governed
by very low temperatures. Thus, reactions involving a barrier can only happen
if they are dominated by tunneling. In the area of biochemistry, however, tunneling is
limited to hydrogen transfers. While the observed KIEs are typically smaller
than in the other fields discussed, they are a valuable probe for the reaction
mechanism and therefore studied extensively. Overall, judging from the dynamic
development in the field of atom tunneling in chemistry in the recent years,
new and exciting findings can be expected for the future as well.

\section{Acknowledgments}
This work was financially supported by the German Research Foundation (DFG)
within the Cluster of Excellence in Simulation Technology (EXC 310/2) at the
University of Stuttgart as well as by the the European Research Council (ERC)
under the European Union’s Horizon 2020 research and innovation programme
(grant agreement No 646717, TUNNELCHEM).  Manuel Weber and Thanja Lamberts are
acknowledged for carefully reading the manuscript.

\providecommand{\url}[1]{\texttt{#1}}
\providecommand{\urlprefix}{}
\providecommand{\foreignlanguage}[2]{#2}
\providecommand{\Capitalize}[1]{\uppercase{#1}}
\providecommand{\capitalize}[1]{\expandafter\Capitalize#1}
\providecommand{\bibliographycite}[1]{\cite{#1}}
\providecommand{\bbland}{and}
\providecommand{\bblchap}{chap.}
\providecommand{\bblchapter}{chapter}
\providecommand{\bbletal}{et~al.}
\providecommand{\bbleditors}{editors}
\providecommand{\bbleds}{eds.}
\providecommand{\bbleditor}{editor}
\providecommand{\bbled}{ed.}
\providecommand{\bbledition}{edition}
\providecommand{\bbledn}{ed.}
\providecommand{\bbleidp}{page}
\providecommand{\bbleidpp}{pages}
\providecommand{\bblerratum}{erratum}
\providecommand{\bblin}{in}
\providecommand{\bblmthesis}{Master's thesis}
\providecommand{\bblno}{no.}
\providecommand{\bblnumber}{number}
\providecommand{\bblof}{of}
\providecommand{\bblpage}{page}
\providecommand{\bblpages}{pages}
\providecommand{\bblp}{p}
\providecommand{\bblphdthesis}{Ph.D. thesis}
\providecommand{\bblpp}{pp}
\providecommand{\bbltechrep}{Tech. Rep.}
\providecommand{\bbltechreport}{Technical Report}
\providecommand{\bblvolume}{volume}
\providecommand{\bblvol}{Vol.}
\providecommand{\bbljan}{January}
\providecommand{\bblfeb}{February}
\providecommand{\bblmar}{March}
\providecommand{\bblapr}{April}
\providecommand{\bblmay}{May}
\providecommand{\bbljun}{June}
\providecommand{\bbljul}{July}
\providecommand{\bblaug}{August}
\providecommand{\bblsep}{September}
\providecommand{\bbloct}{October}
\providecommand{\bblnov}{November}
\providecommand{\bbldec}{December}
\providecommand{\bblfirst}{First}
\providecommand{\bblfirsto}{1st}
\providecommand{\bblsecond}{Second}
\providecommand{\bblsecondo}{2nd}
\providecommand{\bblthird}{Third}
\providecommand{\bblthirdo}{3rd}
\providecommand{\bblfourth}{Fourth}
\providecommand{\bblfourtho}{4th}
\providecommand{\bblfifth}{Fifth}
\providecommand{\bblfiftho}{5th}
\providecommand{\bblst}{st}
\providecommand{\bblnd}{nd}
\providecommand{\bblrd}{rd}
\providecommand{\bblth}{th}

\clearpage
\section*{TOC graphic}

The quantum mechanical tunnel effect is increasingly found to influence many
chemical reactions. While atom-tunneling can only be detected indirectly in
experiment, computational investigations allow a direct observation. Here, we
review cases in which the tunnel effect changes reaction paths and branching
ratios, enables chemical reactions in an astrochemical environment and
influences biochemical processes.

\includegraphics[width=5.5cm]{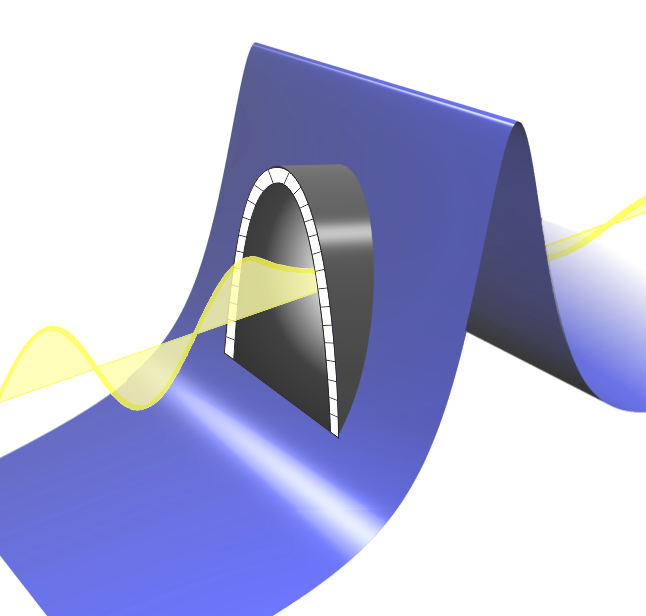}

\end{document}